\documentclass[twocolumn]{aastex62}

\usepackage{amsmath}
\usepackage{multirow}

\shorttitle{Candidate Atmospheres}
\shortauthors{Koll et al.}

\begin{document}

\title{Identifying candidate atmospheres on rocky M dwarf planets via eclipse photometry}

\correspondingauthor{Daniel D.B. Koll}
\email{dkoll@mit.edu}

\author{Daniel D.B. Koll}
\affil{Department of Earth, Atmospheric, and Planetary Sciences, Massachusetts Institute of Technology, Cambridge, MA 02139, USA}

\author{Matej Malik}
\affil{Department of Astronomy, University of Maryland, College Park, MD 20742, USA}

\author{Megan Mansfield}
\affil{Department of Geophysical Sciences, University of Chicago, Chicago, IL 60637, USA}

\author{Eliza M.-R. Kempton}
\affil{Department of Astronomy, University of Maryland, College Park, MD 20742, USA}
\affil{Department of Physics, Grinnell College, 1116 8th Avenue, Grinnell, IA 50112, USA}

\author{Edwin Kite}
\affil{Department of Geophysical Sciences, University of Chicago, Chicago, IL 60637, USA}

\author{Dorian Abbot}
\affil{Department of Geophysical Sciences, University of Chicago, Chicago, IL 60637, USA}

\author{Jacob L.\ Bean}
\affil{Department of Astronomy \& Astrophysics, University of Chicago, Chicago, IL 60637, USA}

%%%%%%%%%%%%%%%%%%%%%%%%%%%%
\begin{abstract}

Most rocky planets in the galaxy orbit a cool host star, {and} there is large uncertainty among theoretical models whether these planets {can} retain an atmosphere.
The \textit{James Webb Space Telescope} (\textit{JWST}) might be able to settle this question empirically, but most proposals for doing so require large observational effort {because they are based on spectroscopy}.
Here we show that infrared photometry of secondary eclipses could quickly identify ``candidate'' atmospheres, by searching for rocky planets with atmospheres thick enough that atmospheric heat transport noticeably reduces their dayside thermal emission compared to that of a bare rock.
{For a planet amenable to atmospheric follow-up}, we find that \textit{JWST} should be able to confidently detect the heat redistribution signal of an $\mathcal{O}(1)$ bar atmosphere with {one to two eclipses}.
{One to two eclipses} is generally much less than the effort needed to infer an atmosphere {via} transmission or emission spectroscopy.
Candidate atmospheres can be further validated via follow-up
spectroscopy or phase curves. {In addition, because this technique is fast it could enable a first atmospheric survey of rocky exoplanets with \textit{JWST}. We estimate that the \textit{TESS} mission will find $\sim100$ planets that are too hot to be habitable but that can be quickly probed via eclipse photometry.}
Knowing whether hot, rocky planets around M dwarfs have atmospheres is important not only for understanding the evolution of uninhabitable worlds: if atmospheres are common on hot planets, then cooler, potentially habitable planets around M dwarfs are also likely to have atmospheres.
\end{abstract}
%%%%%%%%%%%%%%%%%%%%%%%%%%%%

% 

%%%%%%%%%%%%%%%%%%%%%%%%%%%%
\section{Introduction} \label{sec:intro}
%%%%%%%%%%%%%%%%%%%%%%%%%%%%
% ----------
\subsection{The challenge of M dwarf planet atmospheres}
% ----------

The ability of rocky planets orbiting M dwarfs to form and retain atmospheres is a major question in the field of exoplanets because of the forthcoming opportunity to observe these worlds for evidence of habitability and life \citep{shields16,NAP25187}. M dwarfs undergo a long pre-main sequence phase that exposes planets that would later be in the nominal liquid water habitable zone to strong irradiation \citep{chabrier2000} as well as high EUV and solar wind fluxes \citep{dong2018}. Even once on the main sequence, M dwarfs can still exhibit strong flaring events \citep[e.g.,][]{davenport12} and the ratio of their high energy luminosity to bolometric luminosity is substantially larger than for Sun-like stars \citep[e.g.,][]{ribas17,peacock18}. Atmospheric escape on planets orbiting M dwarfs could therefore be extremely high and sustained, raising the possibility that the worlds orbiting in these stars' habitable zones might be predominantly bare rocks with little chance of hosting a surface biosphere \citep[][and references therein]{zahnle17}.

Although volatile loss could be prevalent on M dwarf planets, there are also reasons to be hopeful about the presence of atmospheres on these worlds. These planets might accumulate massive atmospheres in the first place \citep[e.g.,][]{ribas16,barnes16}, could have magnetic fields that would guard against some loss mechanisms \citep[e.g.,][]{segura10}, could outgas secondary atmospheres from their interiors, could have atmospheres with high mean molecular weight gases and thermospheric coolants that suppress atmospheric escape, or could be resupplied with volatiles from an external source, such as through cometary bombardment. The relative efficiency of these processes remains highly uncertain, however, so the final say on whether atmospheres are common on rocky planets around M dwarfs will have to be obtained empirically.

% ----------
\subsection{Current techniques for detecting exoplanet atmospheres}
% ----------

The three main techniques for detecting atmospheres on exoplanets are transmission spectroscopy during a planet's transit (transit spectroscopy), emission spectroscopy during a planet's secondary eclipse (eclipse spectroscopy), and thermal phase curves over the course of a planet's orbit.
Transit and eclipse spectroscopy have been discussed extensively elsewhere \citep{miller-ricci2009,bean10,barstow16,morley2017,louie2018,batalha2018,lustig-yaeger19}. These techniques rely on inferring the spectral signature of atmospheric gases. For a transit that means ruling out a flat transmission spectrum.
For an eclipse that means ruling out a blackbody spectrum and detecting spectral features that are consistent with gas phase {molecules in the planet's atmosphere (although in practice the interpretation can be subtle, see Section 3).}
Thermal phase curves as a means of detecting atmospheres were proposed by \citet{seager2009b}. This technique relies on the signature of an atmosphere's heat redistribution. As long as the planet can be assumed to be tidally locked into synchronous rotation with permanent day- and nightsides, a bare rock planet would exhibit a large day-night temperature
difference in its thermal phase curve, whereas an atmosphere would tend to reduce this temperature difference \citep{seager2009b,selsis2011,koll2016,kreidberg16}.

Unfortunately, all three techniques will likely require substantial investments of observing time \citep[][]{kaltenegger09,deming2009,rauer11,snellen13,rodler14,serindag19}. For transit and eclipse spectroscopy, estimates suggest that atmospheric detection will require anywhere from multiple to more than a dozen repeat observations with the \textit{James Webb Space Telescope} (\textit{JWST}) \citep{batalha2018,louie2018,morley2017}.
Phase curves are inherently costly, because they need to span at least half a planet's orbit, and in some cases the observation might have to be repeated to attain the desired signal-to-noise, thus also requiring long observation periods.

Finally, observations would ideally not just detect the presence of an atmosphere but characterize it in detail. Doing so will be even more expensive than the above estimates suggest because any single technique suffers from a number of degeneracies and false positive scenarios.
For example, transit spectroscopy can be limited by the presence of hazes and clouds \citep[e.g.,][]{kreidberg2014a,sing16}; determining composition from eclipse spectroscopy requires simultaneously determining the atmosphere's thermal structure \citep[e.g.,][]{madhusudhan09,line16a}; and inferring an atmosphere's thickness from its thermal phase curve requires simultaneous knowledge about its composition \citep{koll2015}.
Any effort to move beyond atmospheric detection to detailed characterization will thus likely have to combine multiple techniques, increasing the observational effort even more.

% ----------
\subsection{Our proposal: detecting candidate atmospheres via eclipse photometry}
% ----------

Given how difficult it is to detect and characterize atmospheres on small exoplanets, a fast screening technique is needed to identify those planets that are most promising for follow-up campaigns.
An efficient test for the presence or absence of an atmosphere will also enable exploration of a larger number of planets than can be studied in detail, which is crucial for developing statistical insight into the formation and evolution of planetary atmospheres.

Here we propose such a test, by considering how the planet's atmospheric heat transport affects its dayside thermal emission, which can be measured through its broadband secondary eclipse depth. Our proposal is similar to that of \citet{seager2009b}, but we focus solely on the observable dayside signature.

The energy budget of the planet's dayside can be written as \citep{burrows2014}
\begin{eqnarray}
  T_{day} & = & T_* \sqrt{\frac{R_*}{d}} (1-\alpha_B)^{1/4} f^{1/4}.
\label{eqn:daybudget}
\end{eqnarray}
Here $T_{day}$ is the observed dayside brightness temperature, $T_*$ is the stellar temperature, $R_*$ is the
stellar radius, $d$ is the planet's semi-major axis, $\alpha_B$ is the
planet's Bond albedo, and $f$ is the so-called heat redistribution
factor. 
There are two limits for $f$:
\begin{equation}
  f =
          \begin{cases}
            2/3 & \text{ instant reradiation} \\
            1/4 & \text{ uniform redistribution}
            \end{cases}
\end{equation}
If a planet has no or a sufficiently thin atmosphere then we consider it effectively a bare
rock and $f \rightarrow 2/3$ \citep{hansen2008a}.
Conversely, if the planet has a thick enough atmosphere that its winds redistribute heat between day- and nightside then $f$ will be reduced.
In the limit in which atmospheric heat transport becomes highly
efficient $f \rightarrow 1/4$.
Our proposal amounts to observing $T_{day}$, to infer whether $f$ is significantly smaller than the bare rock limit.

The main promise of atmospheric detection via eclipse photometry is that it does not require high spectral resolution, so it should require less observation time than spectroscopy or phase curves.
Of course, like every other technique, eclipse photometry also suffers from false negatives and false positives.
For example, Equation \ref{eqn:daybudget} shows a degeneracy between $f$ and the albedo $\alpha_B$. Physically, this means a bare rock with high albedo can mimic the dayside thermal emission of a low-albedo planet with a thick atmosphere.
As we discuss in detail in Section \ref{sec:discussion}, we do not believe that this and other degeneracies will greatly affect our proposal, based on both physical modeling and the empirical observation that bare rocks in the Solar System have low albedos.

Nevertheless, because false positives are possible and in analogy to the \textit{Kepler} and \textit{TESS} missions, we consider planets whose dayside brightness temperature strongly deviates from that of a bare rock as ``candidate atmospheres'', but whose atmospheric nature should be confirmed via follow-up.

% ----------
\subsection{Layout of this paper}
% ----------

The goal for the rest of this paper is to quantify how much time is required to infer an atmosphere via eclipse photometry, {how} this effort compares to the effort needed with other atmospheric detection techniques, {and how many planets exist that could potentially be studied with this technique}.
To do so we use atmospheric models to simulate the atmospheres of three nearby rocky planets that are among the best known targets for atmospheric characterization: TRAPPIST-1b \citep{gillon2016,delrez2018}, GJ1132b \citep{berta-thompson2015}, and LHS3844b \citep{vanderspek2019}. Common to all three is that they are too hot to be habitable, which makes them easier to characterize than habitable-zone planets and also decreases the likelihood of false positives for our proposed technique (see discussion).

Our models are described in Section \ref{sec:methods}. We use these models to generate simulated \textit{JWST} observations and compare our results against previous work in Section \ref{sec:jwst_obs}.
We then quantify the observational effort required for detecting an atmosphere using a wide range of techniques, which we present in Section \ref{sec:results}.
For all three modeled planets, we find that a single eclipse observation with \textit{JWST} will be able to confidently detect the atmospheric heat redistribution signal of a thick atmosphere. In contrast, most other techniques will require more observation time. Eclipse photometry is therefore a quick and viable way of inferring atmospheres on rocky exoplanets.
{In Section \ref{sec:future} we then estimate how many other rocky planets exist for which eclipse photometry might be feasible. We find that \textit{TESS} should detect more than 100 rocky planets that this technique could be applied to, which opens up the possibility of a future statistical survey of atmospheres on rocky exoplanets.}
We discuss our results in Section \ref{sec:discussion} and conclude in Section \ref{sec:conclusions}.

%%%%%%%%%%%%%%%%%%%%%%%%%%%%
\section{Methods}
\label{sec:methods}
%%%%%%%%%%%%%%%%%%%%%%%%%%%%

\begin{deluxetable*}{lccccccccc}
\tablecaption{Stellar and planetary parameters.}
\tablecolumns{10}
\tablehead{
\colhead{} & \colhead{R$_*$ (R$_\Sun$)} & \colhead{T$_*$ (K)} & \colhead{R$_p$ (R$_\Earth$)} & \colhead{g (m/s$^2$)} & \colhead{T$_{eq}$ (K)\tablenotemark{a}} &
\colhead{f$_{\mathrm{CO}_2}$\tablenotemark{b}} & 
\colhead{f$_{\mathrm{H}_2\mathrm{O}}$\tablenotemark{b}} &
\colhead{f$_{\mathrm{CO}_2}$\tablenotemark{c}} & \colhead{f$_{\mathrm{H}_2\mathrm{O}}$\tablenotemark{c}}
}
\startdata
TRAPPIST-1b & 0.121 & 2511 & 1.12 & 7.95 & 391 & 0.40 & 0.28 & 0.66 & 0.62 \\
GJ1132b & 0.207 & 3270 & 1.16 & 11.8 & 578 & 0.44 & 0.31 & 0.66 & 0.64 \\
LHS3844b & 0.189 & 3036 & 1.32 & 12.9\tablenotemark{d} & 805 & 0.47 & 0.36 & 0.66 & 0.64
\enddata
\label{tab:parameters}
\tablenotetext{a}{Equilibrium temperature, which assumes full heat redistribution and zero albedo.}
\tablenotetext{b}{Heat redistribution factor, for 1 bar surface pressure.}
\tablenotetext{c}{Heat redistribution factor, for 0.01 bar surface pressure.}
\tablenotetext{d}{Assuming 2.3 $M_{\Earth}$, based on \citet{chen2017a}.}
\end{deluxetable*}

Table \ref{tab:parameters} shows the planetary and host star parameters we use in our calculations for TRAPPIST-1b, GJ1132b, and LHS3844b.
These three planets span a wide range of parameter space, and therefore also function as archetypes for other rocky planets that will be discovered in the near future.
TRAPPIST-1b is the coolest planet we consider with a zero-albedo equilibrium temperature of 391 K. Combined with its low surface gravity, TRAPPIST-1b suggests itself as a target that is most accessible via transit spectroscopy.
GJ1132b has a higher equilibrium temperature of about 578 K, and also a high surface gravity, which tends to favor eclipse spectroscopy. LHS3844b is comparatively hot at 805 K, and also has a short orbital period of just over 11 hours, which makes it a favorable target for both eclipse spectroscopy and thermal phase curves.

\begin{figure*}
\includegraphics[width=\linewidth, clip]{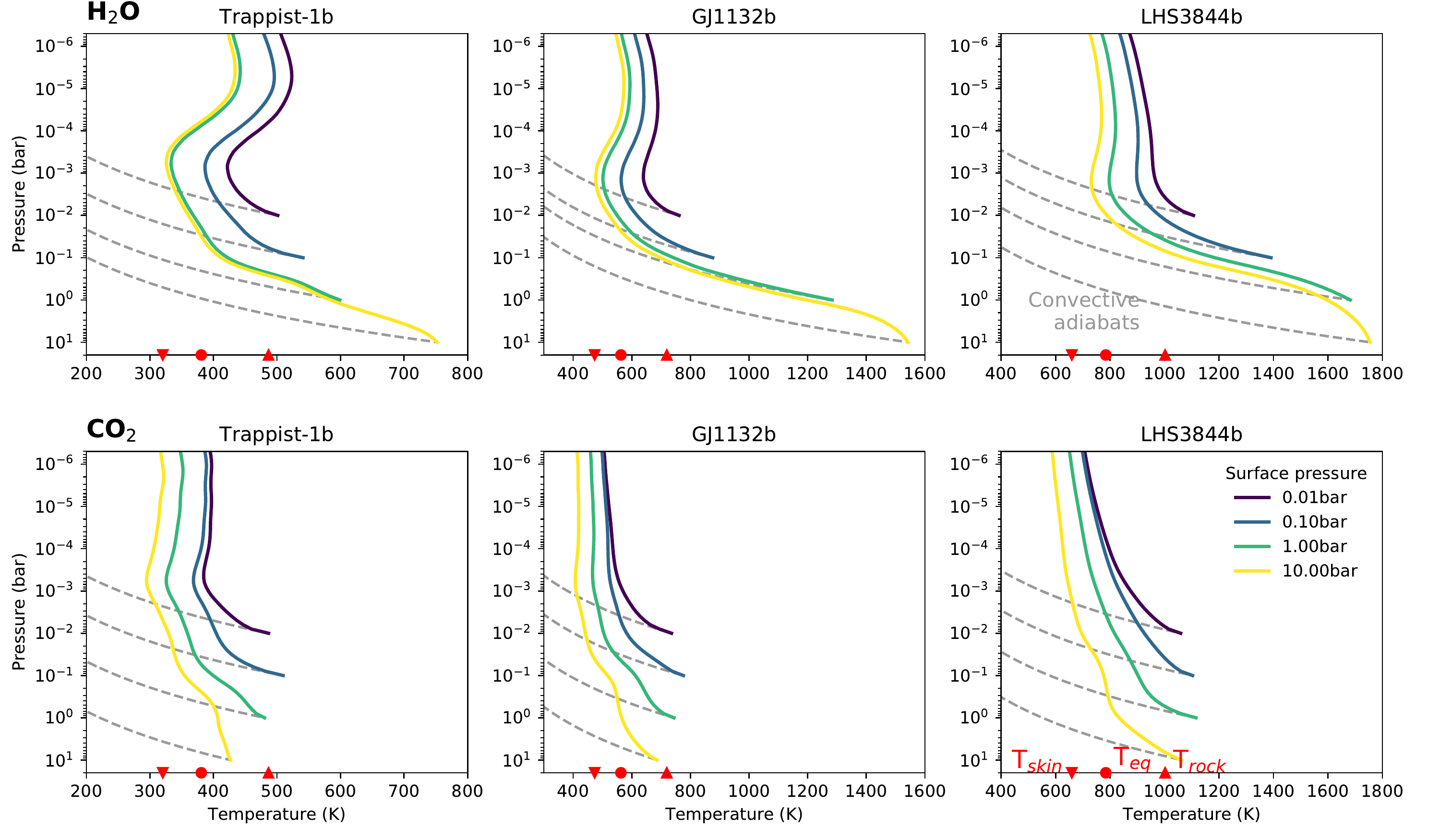}

\caption{Dayside temperature-pressure profiles as a
  function of surface pressure. Each row corresponds to a different
  atmospheric composition, each column corresponds to a different
  planet. Dashed grey lines show convective adiabats. Vertical
  temperature profiles are generally less steep than adiabatic, and are largely set by radiative transfer. Red symbols at the bottom show theoretical
  limits: $T_{rock}$ = emission temperature of a bare rock, which corresponds to no heat redistribution (right side up triangle), $T_{eq}$ = equilibrium temperature, which corresponds to full heat redistribution (circle), and $T_{skin}$ = skin temperature of a grey stratosphere, which is equal to $T_{eq}/2^{1/4}$ (upside down triangle).
  }
\label{fig:TPprofiles}
\end{figure*}

For each planet we simulate a range of different atmospheric scenarios.
We consider eight scenarios that cover four different surface pressures, ranging from 10$^{-2}$ to 10 bar, and two different atmospheric composition end-members, namely pure H$_2$O (steam) and pure CO$_2$.

For each of these scenarios we use a 1D atmospheric column model, \texttt{HELIOS}, to simulate the dayside-averaged temperature-pressure (T-P) profile and the planet's emission spectrum. We also use \texttt{HELIOS} to simulate the T-P profile near the terminator, which we use as input to compute the planet's transmission spectrum with a second model, \texttt{Exo-Transmit}. Below we describe our models in detail.

\texttt{HELIOS} is a 1D column model that uses hemispheric two-stream radiation and convective adjustment to simulate a dayside-averaged atmosphere in radiative-convective equilibrium \citep[][Malik et al.\ submitted]{malik2017,malik19}.
We do not include condensation, so convection adjusts the atmosphere in unstable layers back to a dry adiabat.
{For the surface we use a spectrally uniform albedo of 0.1, where the chosen value is motivated by a companion paper in which we consider the potential albedos of rocky exoplanet surfaces in more detail (Mansfield et al., submitted).}
For the radiative transfer we use ExoMol line lists for H$_2$O \citep{barber2006} and HITEMP for CO$_2$ \citep{rothman2010}, calculated with \texttt{HELIOS-K} \citep{grimm2015}. We approximate the spectral lines with a Voigt profile and sub-Lorentzian wing cut-off at 100 cm$^{-1}$ from line center. Pressure broadening is included using the default broadening parameters from the ExoMol webpage and the self-broadening parameters from the HITRAN/HITEMP database. Further included is CO$_2$-CO$_2$ collision-induced absorption \citep{richard2012}, and Rayleigh scattering of H$_2$O and CO$_2$ \citep{cox2000, sneep2005, wagner2008, thalmann2014}. The radiative transfer calculation is performed using 300 wavelength bins between 0.33 and 1,000 $\mu$m, employing the correlated-$\kappa$ assumption with 20 Gaussian points in each bin. The final emission spectra are post-processed at a resolution of R=3000.

Because \texttt{HELIOS} is a vertical 1D model, it cannot resolve the atmosphere's horizontal heat redistribution between day- and nightside. We parameterize the heat redistribution as a function of surface pressure and atmospheric composition with a theoretical scaling that is derived in another companion paper (Koll, submitted). Briefly, the scaling parameterizes the heat redistribution factor $f$ as
\begin{eqnarray}
 f & = & \frac{2}{3} - \frac{5}{12} \times
                        \frac{\tau_{LW} \left(\frac{p_s}{1
                        \mathrm{bar}}\right)^{2/3}\left(\frac{T_{eq}}{600\mathrm{K}}\right)^{-4/3}}{k
                        +\tau_{LW} \left(\frac{p_s}{1
                        \mathrm{bar}}\right)^{2/3}\left(\frac{T_{eq}}{600\mathrm{K}}\right)^{-4/3}}.
\end{eqnarray}
Here $p_s$ is the surface pressure, $T_{eq}$ is the planet's equilibrium temperature, $k \approx 2$, and $\tau_{LW}$ is the broadband longwave optical thickness. {The heat redistribution factor correctly reduces to $f=1/4$ for a thick atmosphere with strong infrared absorption ($p_s,\tau_{LW}$ become large) and $f=2/3$ for a vanishingly thin atmosphere ($p_s,\tau_{LW}\rightarrow 0$). We define the} broadband optical thickness $\tau_{LW}$ for a given atmospheric composition and surface pressure based on the atmosphere's attenuation of the surface's thermal emission,
\begin{eqnarray}
  \tau_{LW} & = & -\ln\left[  \frac{\int e^{-\tau_\lambda}
                  B_\lambda(T_s)d\lambda}{\int
                  B_\lambda(T_s)d\lambda} \right].
\end{eqnarray}
Here $B_\lambda$ is the Planck function, $T_s$ is the surface temperature, and $\tau_\lambda$ is the atmosphere's column-integrated optical {depth} at a given wavelength computed with \texttt{HELIOS}. {Table \ref{tab:parameters} shows values of $f$ in our simulations with 1 bar and 0.01 bar surface pressure. Atmospheres with 1 bar surface pressure have a heat redistribution that clearly deviates from a bare rock, whereas in thinner atmospheres heat redistribution becomes inefficient.}
For our transmission spectra we use \texttt{Exo-Transmit} \citep{kempton2017a}. We use the standard opacity data tables included with \texttt{Exo-Transmit} for 100\% H$_2$O and 100\% CO$_2$ atmospheres. As input we use T-P profiles generated from \texttt{HELIOS}, which differ from the ones we use to generate emission spectra only in that they are calculated at a zenith angle of 80 degrees, appropriate for regions near the planet's limb. 

For our \textit{JWST} noise calculations we use \texttt{PandExo} \citep{batalha2017}.
We use a saturation limit of 50\% full well to avoid a potentially nonlinear detector response at higher electron counts,
and assume an out-of-transit baseline that is 4 times as long as the transit duration. We allow \texttt{PandExo} to optimize the number of groups per integration. We do not include an inherent noise floor, which is an optimistic assumption.
We do so because it allows us to better compare our results with previous studies which also did not include a noise floor \citep{morley2017,batalha2018}.

For reference, \citet{greene2016} suggested a noise floor of 20 ppm for a single observation with NIRSpec and 50 ppm for a single observation with MIRI. We find that roughly half of our estimated errors for TRAPPIST-1b and LHS3844b fall below these thresholds, while for GJ1132b almost all errors fall below these thresholds due to its brighter host star. Even though our noise calculations are thus optimistic and could be affected by systematics, we also find that almost all errors lie within a factor of 2 of the thresholds suggested by \citet{greene2016}. As a conservative estimate, systematics could thus increase our observation times in Section \ref{sec:results} by at most a factor $2^2=4$, with planets around bright host stars such as GJ1132b most likely to be affected.
These values are highly uncertain, however, and \textit{JWST}'s actual performance remains to be seen. The results from the Transiting Exoplanet Community Early Release Science Program will help bring clarity to this issue \citep{bean18}.

For the host stars we use blackbody spectra in our emission calculations, and spectra from the \texttt{PHOENIX} online library \citep{husser2013} for the noise calculations with \texttt{PandExo}. We interpolate the \texttt{PHOENIX} spectra for the stellar temperatures in Table~\ref{tab:parameters} and additionally use $\log_{10} g_{\rm star} = 5.22^y, 5.06^y, 5.06$, and [M/H] = 0.04, -0.12, 0 for TRAPPIST-1b \citep{delrez2018}, GJ 1132 \citep{berta-thompson2015} and LHS 3844 \citep{vanderspek2019}, respectively. The $\log_{10} g_{\rm star}$ values marked with a $y$ are our own estimates, which we derive from the stellar mass. The metallicity of LHS 3844 is unknown, which is why we adopt solar metallicity for that star.

Figure \ref{fig:TPprofiles} shows the dayside-average temperature profiles that we simulate with \texttt{HELIOS}. We find that stratospheric inversions are common, particularly for H$_2$O atmospheres. Because we do not include other absorbers here, such as TiO, the inversions have a different cause than those on hot Jupiters. {These inversions also occur if we use \texttt{PHOENIX} spectra instead of blackbodies for the host stars}. Instead, the inversions are caused by strong {atmospheric} absorption in the near-IR together with {the cool host stars emitting most strongly in the infrared},
%the very red spectra of the host stars,
which we explain in detail in a third companion paper (Malik et al.\ submitted).

Figure \ref{fig:TPprofiles} also shows that radiation is generally more influential than convection in setting vertical temperature structures. The dashed lines in Figure \ref{fig:TPprofiles} show adiabatic profiles for comparison. In a few cases the lowest scale height is close to an adiabat, such as for an H$_2$O atmosphere on GJ1132b and LHS3844b, but in most cases convection is either confined to a narrow surface layer or altogether absent (e.g., TRAPPIST-1b with 10 bars of CO$_2$).

%%%%%%%%%%%%%%%%%%%%%%%%%%%%
\section{Simulated observations with \textit{JWST}}
\label{sec:jwst_obs}
%%%%%%%%%%%%%%%%%%%%%%%%%%%%

% ----------
\subsection{Simulated observations}
% ----------

We process our simulations to wavelength ranges that will be observable with \textit{JWST}. For transit spectroscopy we consider NIRSpec/G235M between 1.66-3.07 $\mu$m. For eclipse spectroscopy we consider MIRI/LRS between 5-12 $\mu$m.

\begin{figure*}
\includegraphics[width=\linewidth, clip]{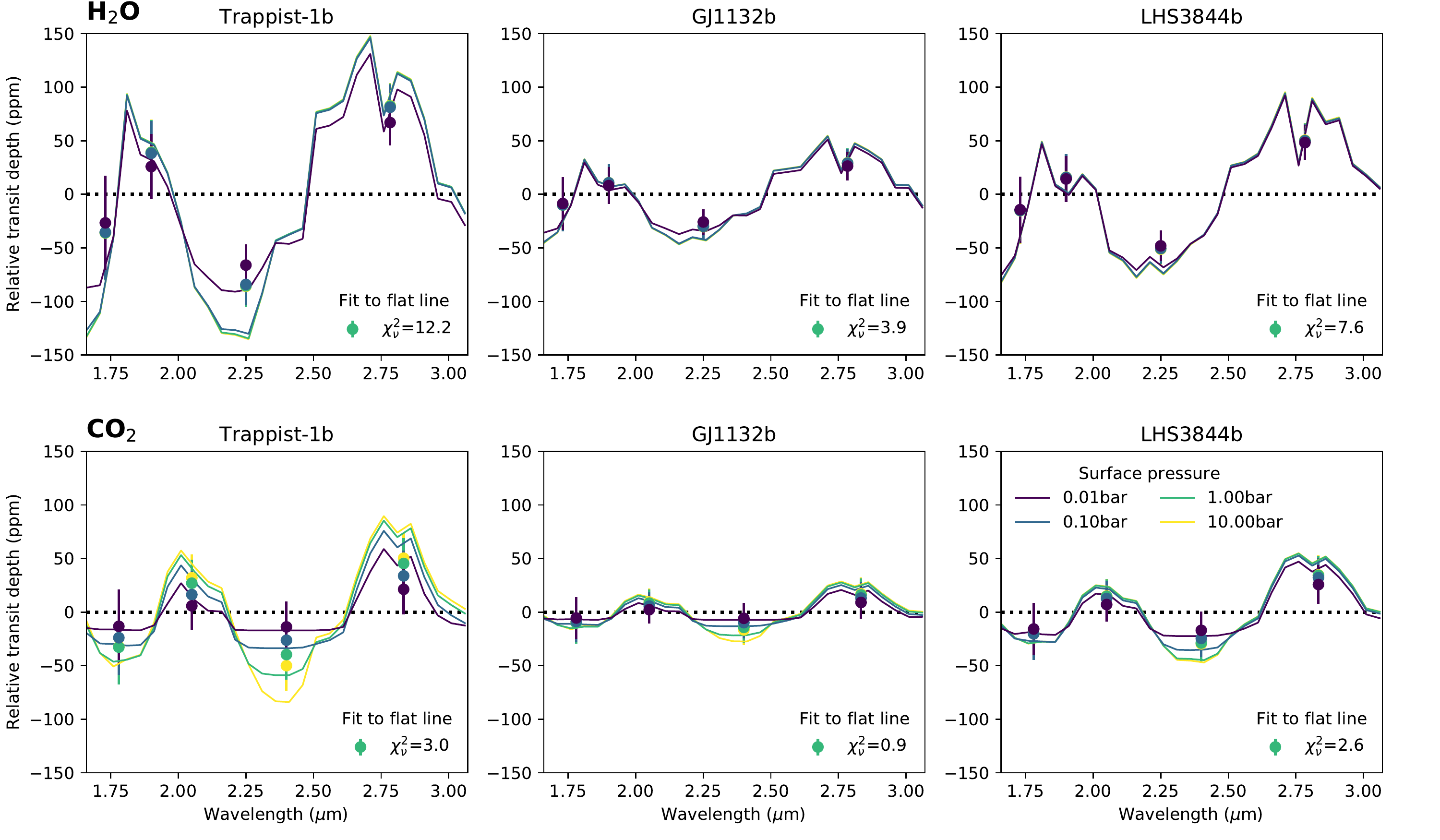}

\caption{Transit spectra observed by \textit{JWST}. Each row corresponds to a different atmospheric
  composition, each column corresponds to a different planet, and colors correspond to different surface pressures. Here
  $\chi_\nu^2$ indicates the goodness of fit to a flat line (black dots). For simplicity we only show $\chi_\nu^2$ for
  simulations with 1 bar surface pressure.}
\label{fig:transit_obs}
\end{figure*}

To evaluate whether \textit{JWST} can detect an atmosphere we use a simple $\chi^2$ metric.
Qualitatively, if the reduced $\chi^2$ value, $\chi_\nu^2=\chi^2/\nu$, is much bigger than $\chi_\nu^2 \sim 1 + \sqrt{2/\nu}$, which is of order unity, we can reject the null hypothesis that the planet is a bare rock and consider this an atmospheric detection.
Here $\nu$ is the number of degrees of freedom in a given spectrum.
For example, with $\nu=3$, we might have some confidence in an atmospheric detection once $\chi_\nu^2 \gg 1 + \sqrt{2/3} = 1.8$. More formally, the probability that $\chi_\nu^2 > 2$ is 11\% and the probability that $\chi_\nu^2 > 3$ is 3\%.
The null hypothesis, and thus the $\chi^2$ value, as well as $\nu$ have to be defined differently for each technique as follows.

For transit spectroscopy we compute $\chi^2$ from the fit between the observed wavelength-dependent transit depth and a flat line, where the flat line is simply the average transit depth in the NIRSpec wavelength range.
Because the flat line is derived from the observations, $\nu$ is equal to the number of observed datapoints minus one.

{For eclipse spectroscopy we require observations to rule out a blackbody to count as an atmospheric detection.
We note that this definition is susceptible to false positives: bare rocks can have spectral features (see discussion) and a planet's emission spectrum can be contaminated by reflected stellar light. For cool stars this could impart molecular features in emission that are due to molecules in the star's, not the planet's, atmosphere.
There are also possible false negatives: an atmosphere with very thick clouds could hypothetically resemble a blackbody. Such an atmosphere would be undetectable via spectroscopy, and detection would instead need to rely on eclipse photometry or thermal phase curves.
Our detection metric for eclipse spectroscopy is therefore over-confident, and in practice atmospheres might be more difficult to detect using this technique. 
}

{The temperature of the null hypothesis blackbody,} which physically corresponds to the planet's dayside brightness temperature, is a priori unknown {so} we use the same two-step procedure as one would follow with actual observations. First, we fit a blackbody to the observed spectrum using \texttt{scipy.optimize.curve\_fit}, optimizing for the blackbody's temperature.
Second, we compute $\chi^2$ from the fit between the observed emission spectrum and the best-fit blackbody spectrum.
Because one degree of freedom is used to fit the blackbody's temperature, $\nu$ is again equal to the number of observed datapoints minus one.

For eclipse photometry we compute $\chi^2$ from the difference between the observed emission spectrum and a blackbody spectrum that assumes no heat redistribution. 
We assume that the surface albedo of the no-heat-redistribution blackbody is known and equal to 0.1 (see Section \ref{sec:discussion}).
Although we are computing a photometric signal, we use the same spectral resolution as for eclipse spectroscopy.
In theory we could bin even further, to a single photometric datapoint, but in practice we find that increased binning leads to little increase in statistical significance for many cases.
Because the eclipse depth of the no-heat-redistribution blackbody is defined independently of any observed datapoints, $\nu$ is equal to the number of observed datapoints.

For phase curves we compute $\chi^2$ from the phase curve amplitude, i.e., the day-night emission difference. To do so we first generate a nightside emission spectrum by rescaling the emitted dayside flux via a spectrally uniform factor that depends on the atmosphere's heat redistribution, $F_{night} = 3/5 \times (2/3-f)/f \times F_{day}$. This expression guarantees the correct nightside fluxes in the thick and thin atmosphere limits.
We then compare the phase amplitude (i.e., the day-night flux difference) of the planet with an atmosphere to the phase amplitude of a bare rock, and set $\nu$ equal to the number of observed datapoints.
We note that phase curves contain additional information that can be used to infer the presence of an atmosphere, such as hot spot offsets. Here we only focus on the phase curve amplitude because global climate models suggest that hot spot offsets become negligible on rocky planets with relatively thin atmospheres \citep{koll2015,koll2016}.

We note that the $\chi^2$ metric is overly conservative because it does not capture spectral correlations. For example, a high-resolution transit spectrum could have a small $\chi^2$ value relative to a flat line, yet still show clear correlation between nearby points that are part of a spectral band.
A full retrieval model would be able to detect this band structure, and thus infer an atmosphere, whereas a simple $\chi^2$ test might miss it.
To account for this effect we downsample all simulated spectra to low spectral resolution, so each spectral point corresponds to a single spectral band. To downsample we weight simulated data by the inverse variance at each wavelength{, so points with smaller error bars contribute more to the spectral mean than points with larger error bars. In practice this mostly affects the MIRI/LRS bandpass, where detector efficiency as well as stellar photon count decrease notably between 5 $\mu$m and 12 $\mu$m. We} select the low-resolution spectral bands by hand for each atmospheric composition. A retrieval algorithm would have to infer these bands from the data, so by giving ourselves this information we are increasing the likelihood of detecting a spectral signature (i.e., a real spectral retrieval would be less confident in detecting an atmosphere than our hand-tailored approach).

Figure \ref{fig:transit_obs} shows what our synthetic \textit{JWST} transit spectra look like, with error bars representing a single transit.
For both H$_2$O and CO$_2$ atmospheres we bin the transit down to just four spectral points, which capture the dominant bands and windows of each gas in the NIRSpec wavelength range.
Figure \ref{fig:transit_obs} also shows the $\chi_\nu^2$ values relative to a flat line; for ease of viewing we only show $\chi_\nu^2$ for the simulations with 1 bar surface pressure.
For H$_2$O atmospheres we find $\chi_\nu > 1$ in all cases. The best transit target is TRAPPIST-1b with $\chi_\nu = 12.2$ but even the worst target, GJ1132b, has $\chi_\nu = 3.9$.
A single transit spectrum should thus be sufficient to infer an atmosphere.
The $\chi_\nu^2$ values are smaller for CO$_2$, due to CO$_2$'s higher mean-molecular-weight (MMW) and smaller scale height, but even here we find $\chi_\nu = 3.0$ for TRAPPIST-1b and $\chi_\nu = 2.6$ for LHS3844b. 
The only scenario in which a single transit is not sufficient to detect an atmosphere is GJ1132b with a CO$_2$ atmosphere for which $\chi_\nu^2 = 0.9$.
We note, however, that the detectability of these transit spectra is optimistic because they do not include any atmospheric aerosols. We explore the possible impact of clouds and hazes on transit spectroscopy in the next Section.

\begin{figure*}
\includegraphics[width=\linewidth, clip]{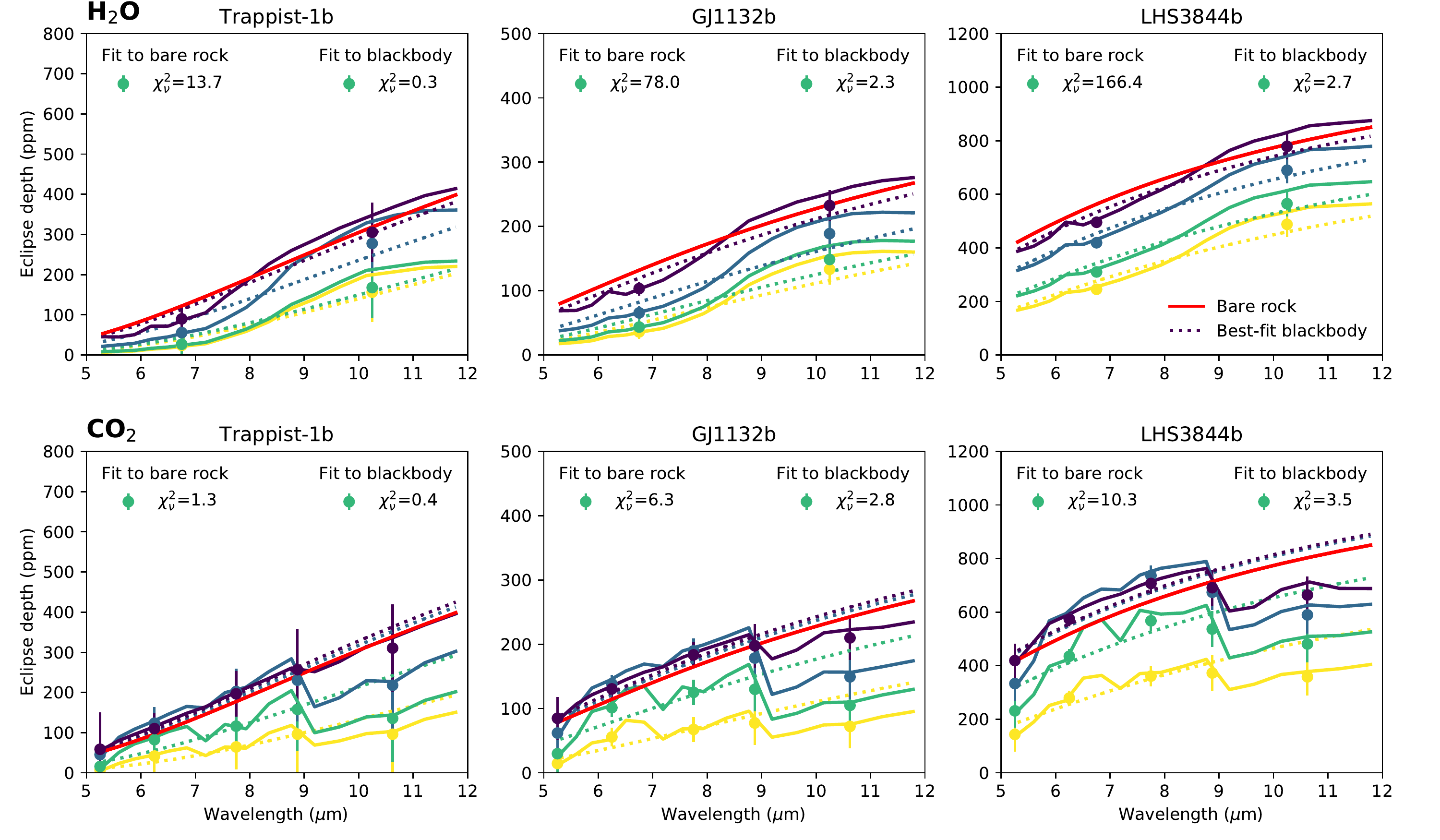}
\caption{Emission spectra observed by \textit{JWST}. Each row corresponds to a different atmospheric
  composition, each column corresponds to a different planet. Here
  $\chi_\nu^2$ indicates the goodness of fit assuming the planet is a bare rock without any heat redistribution (red line), and the goodness of fit assuming the planet is a blackbody (dotted lines). 
  Colors correspond to the same surface pressures as in Figure \ref{fig:transit_obs}, and range from 0.01 bar (purple) to 10 bar (yellow). For simplicity we only show $\chi_\nu^2$ for simulations
  with 1 bar surface pressure. Note the different y-limits for different planets.}
\label{fig:eclipse_obs}
\end{figure*}

Figure \ref{fig:eclipse_obs} shows what our synthetic \textit{JWST} emission spectra look like, with error bars representing a single eclipse.
For H$_2$O we bin the data down to just two spectral points in the MIRI/LRS bandpass, for CO$_2$ we use five spectral points.
Figure \ref{fig:eclipse_obs} shows the $\chi_\nu^2$ values for an observed emission spectrum relative to a best-fit blackbody and relative to a bare rock blackbody.
We find that it is difficult to detect spectral features via eclipse spectroscopy for cool planets, while warm planets are feasible targets. For example, we find $\chi_\nu^2 = 0.3$ for TRAPPIST-1b with an H$_2$O atmosphere whereas $\chi_\nu^2 = 3.5$ for LHS3844b with a CO$_2$ atmosphere.

In contrast to eclipse spectroscopy, we find that eclipse photometry can detect atmospheres with heat redistribution {in the vast majority of cases}. Even for a cool planet like TRAPPIST-1b with a H$_2$O atmosphere, we find that 1 bar of atmosphere leads to a notable difference between the dayside's broadband emission and a bare rock's ($\chi_\nu = 13.7$). {The only case in which a single eclipse is not sufficient for a confident detection is TRAPPIST-1b with a CO$_2$ atmosphere, for which $\chi_\nu = 1.3$.} {The ease of detection via eclipse photometry strongly} increases with temperature, and LHS3844b with 1 bar of CO$_2$ deviates very strongly from a bare rock ($\chi_\nu = 10.3$).

\begin{figure*}
\centering
\includegraphics[width=0.85\linewidth, clip]{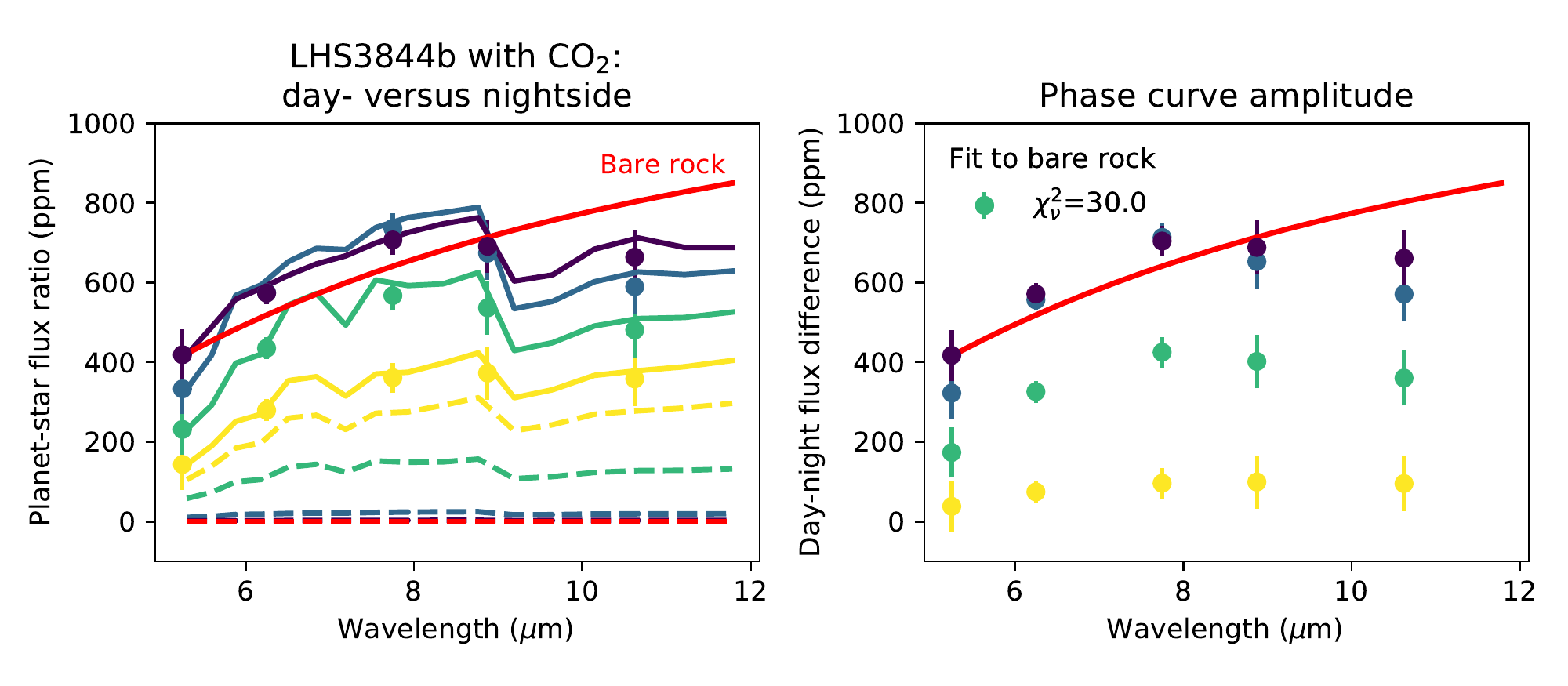}
\caption{Illustration of how we compute phase curve amplitudes. Shown is LHS3844b with a CO$_2$ atmosphere. Left: Solid lines show the simulated dayside spectra, dashed lines show the inferred nightside spectra. Right: the phase curve amplitude is equal to the day-night flux difference, and $\chi^2$ is relative to the phase curve amplitude of a bare rock. Colors correspond to the same surface pressures as in Figure \ref{fig:transit_obs}, and range from 0.01 bar (purple) to 10 bar (yellow).}
\label{fig:phasecurve}
\end{figure*}

Figure \ref{fig:phasecurve} illustrates how we compute the \textit{JWST} phase curve signal. Here we show LHS3844b with a CO$_2$ atmosphere. First, we rescale the simulated dayside to get a nightside emission spectrum. We then compute the day-night flux difference, and compare this difference to the day-night difference of a bare rock.
The error bars are the same as in Figure \ref{fig:eclipse_obs}, which amounts to binning the observed phase curve into bins of 31 min (the duration of a transit or eclipse for LHS3844b).
We find that, similar to eclipse photometry, phase curves should be able to infer thick atmospheres with high confidence {and a 1 bar atmosphere on LHS3844b would be ruled out with $\chi_\nu = 30$}.
The high confidence of this atmospheric detection, however, has to be weighed against its (potentially high) observational cost, which we consider in the next section.

% ----------
\subsection{Comparison with previous work}
% ----------

Our results qualitatively agree with previous \textit{JWST} calculations, even though we employ a number of different modeling assumptions and we simulate different instrument modes.

\citet{batalha2018} computed signal-to-noise (SNR) for transit observations of cool habitable-zone planets and found that about 10 repeated transits with the NIRSpec Prism mode are needed to detect spectral features. This number is much larger than the single transit we find here, but \citet{batalha2018} focused on cooler planets and included high-altitude clouds in their calculations. Indeed, as we show in the next section, clouds and hazes greatly increase the observational effort to detect an atmosphere via transit spectroscopy.

\citet{louie2018} computed SNR for transit observations of warm super-Earths with H$_2$O atmospheres using the NIRISS instrument and found that 10h of telescope time (about 2-3 transits) are sufficient to detect spectral features on TRAPPIST-1b and GJ1132b with a high SNR of about 20-40. These numbers suggest that a single transit would be sufficient for atmospheric detection, in agreement with our results. Moreover, our $\chi^2$ calculation is similar to their SNR metric, and we find that the two metrics agree to within a factor of 3 or better once we account for the different observation lengths.

{
\citet{molliere2016} simulated transit spectra of GJ 1214b with a cloudy, relatively high mean molecular weight atmosphere and found that about 10 transits with NIRSpec could rule out a flat line with 95\% probability. Although their result is again strongly affected by clouds and hazes and considered a different planet, it is to order of magnitude comparable with estimates we present in the next section for how clouds and hazes can impact transmission spectroscopy.
}

\citet{morley2017} estimated the amount of time required to characterize an atmosphere for both transit and eclipse spectroscopy using the same set of instrument modes as we do. They found that for a favorable transit target like TRAPPIST-1b with a CO$_2$ atmosphere, about six transits are needed to rule out a flat line at 5$\sigma$ confidence, while other targets would require longer observations.
Similarly, for a favorable emission target like GJ1132b, about 2-3 eclipses are needed to detect the secondary eclipse at 25$\sigma$ confidence.
We will show in the next section that these results are comparable to our own calculations, even though our estimates are slightly more optimistic. For example, we estimate that a 5$\sigma$ detection of CO$_2$ on TRAPPIST-1b will require about four transits, compared to \citeauthor{morley2017}'s six transits.

We note that even though our detectability estimates are comparable to those of previous groups, we use different physical assumptions.
For example, we simulate temperature profiles in self-consistent radiative-convective equilibrium. In contrast, \citet{batalha2018} assumed a parameterized temperature profile that was based on analytic grey calculations, and \citet{morley2017} assumed that all atmospheres are convective up to a pressure of 0.1 bar and are capped by an isothermal stratosphere whose temperature is equal to the skin temperature.

The different assumptions about temperature profiles should only have a small effect on transit spectra, but they will affect emission spectra.
We find that convection is generally suppressed due to the red host star spectra and atmospheric shortwave absorption (Fig.~\ref{fig:TPprofiles}). 

The clear majority of our simulations also do not show a transition between convective and radiative zones at around 0.1 bar, which has been proposed based on Solar System atmospheres \citep{robinson2014b}.
This means vertical temperature gradients on M dwarf planets should generally be smaller, and signals for emission spectra lower, than one might predict with parameterized convective temperature profiles.
%

%%%%%%%%%%%%%%%%%%%%%%%%%%%%
\section{Comparing detection efficiency for different observation strategies} \label{sec:results}
%%%%%%%%%%%%%%%%%%%%%%%%%%%%

\begin{figure*}
\includegraphics[width=\linewidth, clip]{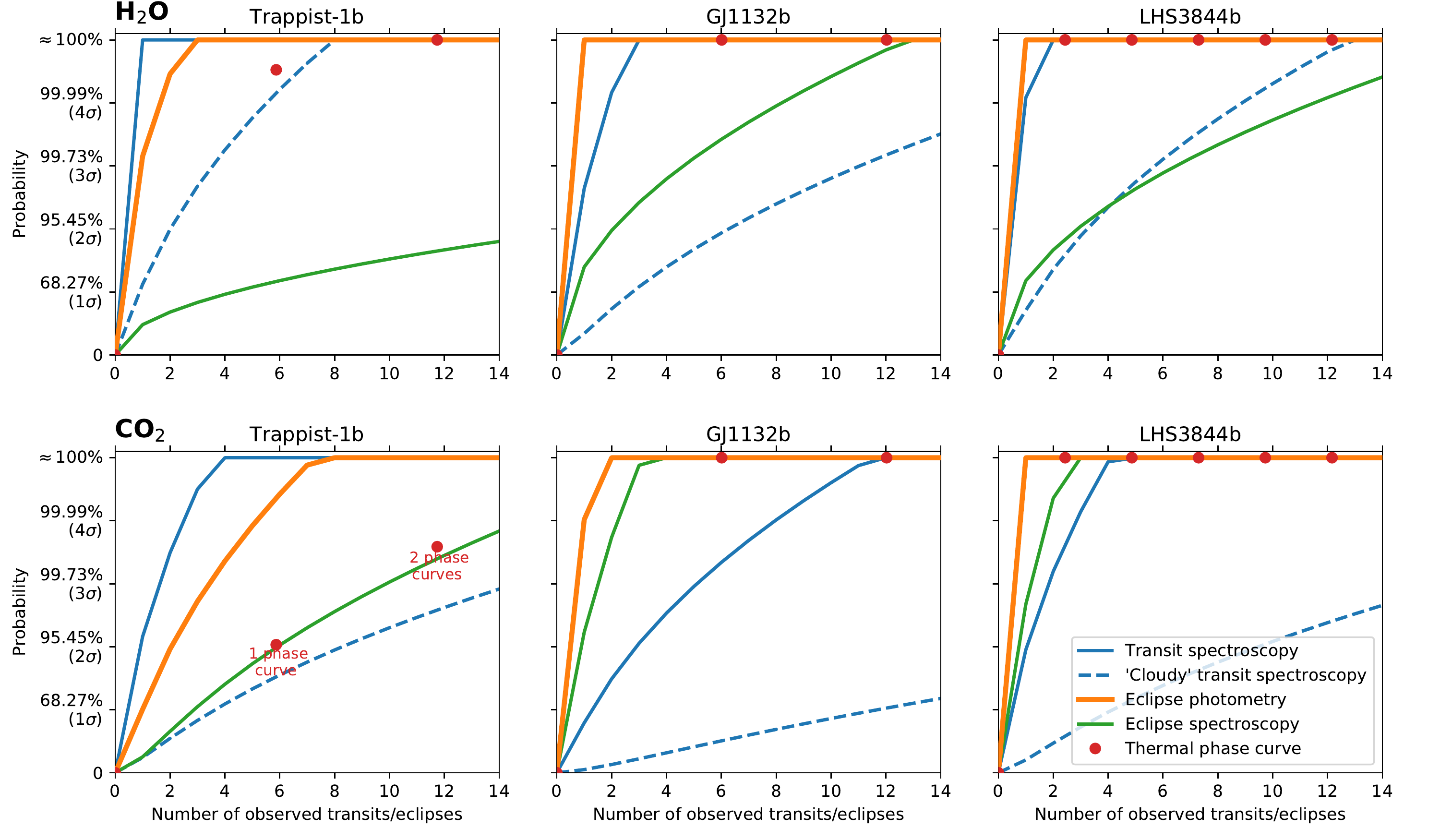}
\caption{This plot shows how many repeated transit or eclipse observations 
  with \textit{JWST} are required to detect the presence of a 1 bar atmosphere. We compare four different detection methods: transit spectroscopy, eclipse spectroscopy, eclipse photometry, and phase curves. For transit spectroscopy, we also include a `cloudy' case with reduced signal amplitude (see Section \ref{sec:results}). For phase curves we convert the observation time needed into an equivalent amount of transits or eclipses. Because we assume 1 bar of surface pressure, the atmosphere is thick enough to significantly modify the planet's heat redistribution.}
\label{fig:bang_per_buck}
\end{figure*}

In this section we combine the results from the previous section to address our initial question: how much telescope time is needed to infer an atmosphere via eclipse photometry, and how does this effort compare to the effort required with other techniques?
To do so we take the $\chi^2$ values we computed for a single transit or eclipse in the previous section, and use the $\chi^2$ distribution to convert them into a probability of ruling out the no-atmosphere null hypothesis.
To compute how the detection probability increases with the number of repeated measurements we assume photon noise and scale our \textit{JWST} error bars for a single transit or eclipse by $1/\sqrt{N_{obs}}$.

We consider two planetary scenarios: an optimistic scenario with a surface pressure of 1 bar, so that the atmosphere is thick enough to substantially affect the day-night heat redistribution as well as the observed spectral features, and a pessimistic scenario with a surface pressure of 0.01 bar, so that the atmosphere's day-night heat redistribution and observable spectral features are weak.
H$_2$O and CO$_2$ atmospheres become increasingly vulnerable to atmospheric collapse at surface pressures lower than 0.1-0.01 bar, so this surface pressure is a reasonable lower bound for the atmospheric compositions we consider here \citep{joshi1997,wordsworth2015,koll2016}.

We compare the four techniques for inferring an atmosphere laid out in Section \ref{sec:jwst_obs}: transit spectroscopy, eclipse spectroscopy, eclipse photometry, and thermal phase curves.
We note that our calculated spectra do not include aerosols, which is an optimistic assumption because high-altitude clouds or hazes can curtail the amplitude of observable spectral features. This assumption is particularly problematic for atmospheric detection via transit spectroscopy, because transit spectroscopy is more sensitive to clouds than other techniques \citep{fortney2005} and because a cloudy transit spectrum cannot be distinguished from a bare rock.
Motivated by observations of hot Jupiter transits, we therefore add a `cloudy' transit scenario in which we multiply the amplitude of our simulated transit spectra by 1/3 \citep{wakeford2019}.

To compare each observation technique in terms of its observational effort, we convert the telescope time necessary to measure a phase curve into an equivalent number of transit or eclipses. Phase curves cover half the planet's orbital period, and we assume that every observation requires 3 hours of additional overhead time due to telescope slew, detector burn-in, and measurement of the star's out-of-transit/eclipse baseline flux, similar to the value used in \citet{louie2018}.
For reference, phase curves of TRAPPIST-1b and GJ1132b require about as much \textit{JWST} time as six transits or eclipses of the same planets.
%

% ----
Figure \ref{fig:bang_per_buck} shows our main result. We display probabilities as well as an approximate detection significance in terms of $\sigma$ confidence levels (e.g., a 95.45\% probability is equal to 2$\sigma$). Once probabilities exceed 5$\sigma$ we round up to 100\%.
The top left of Figure \ref{fig:bang_per_buck} represents cool atmospheres with large scale heights, while the bottom right of Figure \ref{fig:bang_per_buck} represents hot atmospheres with small scale heights. As we discussed in Section \ref{sec:jwst_obs}, signals are large for transit spectroscopy in the top left while signals are large for eclipse spectroscopy in the bottom right.

Figure \ref{fig:bang_per_buck} shows that eclipse photometry is very promising for atmospheric detection. As long as the atmosphere is thick enough to induce a significant heat redistribution, eclipse photometry should be able to detect this signal with {one to two} \textit{JWST} eclipses.
For {most} planet scenarios shown in Figure \ref{fig:bang_per_buck} we find that a single \textit{JWST} eclipse should be able to rule out a bare rock with more than {3$\sigma$} confidence. {The only exception is Trappist-1b with a CO$_2$ atmosphere, for which two eclipses are required to reach 2$\sigma$. In all other cases two eclipses are sufficient to rule out a bare rock at more than 4$\sigma$}.
Eclipse photometry could therefore be used to quickly search {favorable} rocky exoplanets for atmospheric signatures, as revealed by their dayside emission temperatures.

\begin{figure*}
\includegraphics[width=\linewidth, clip]{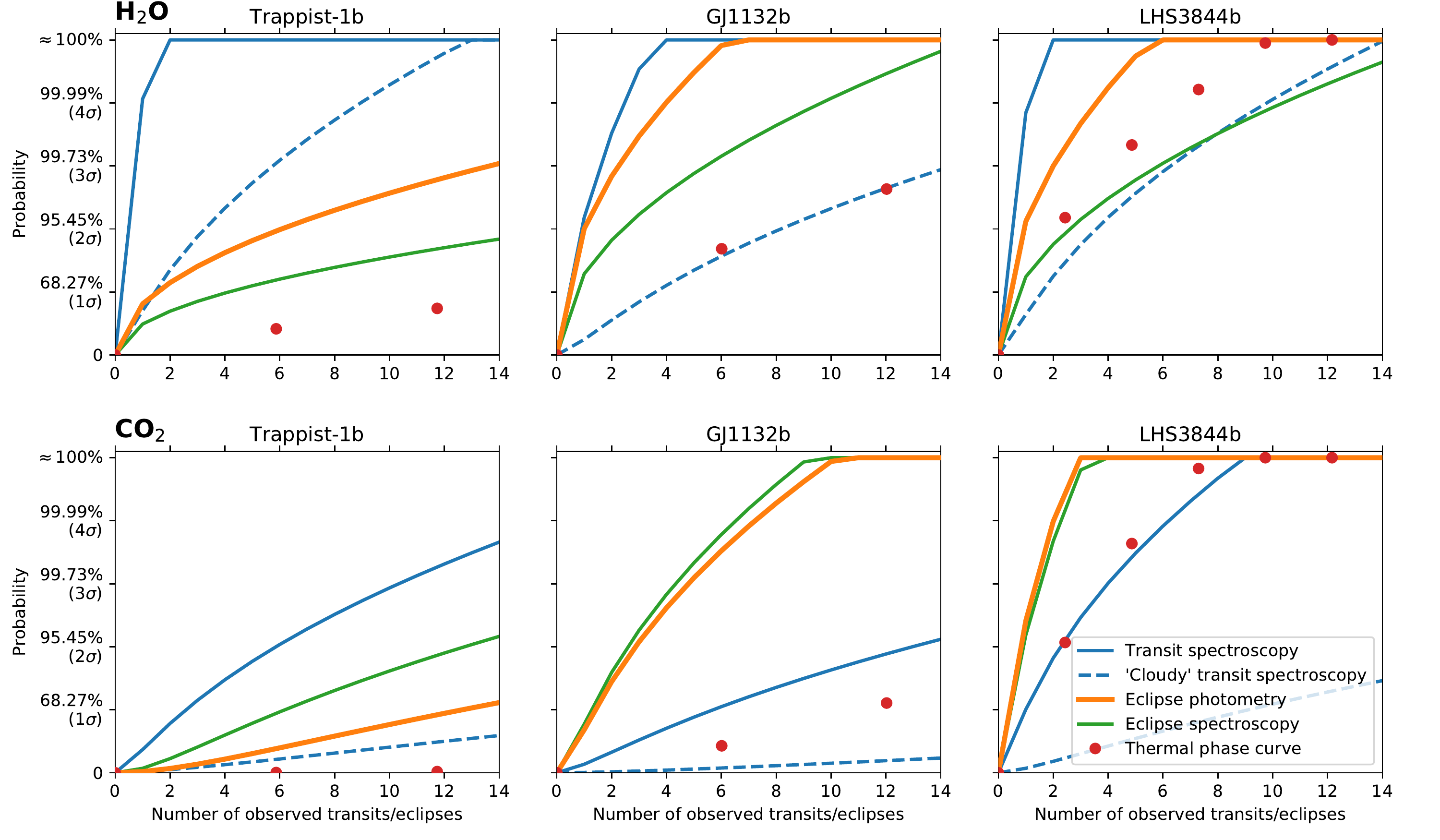}
\caption{Similar to Figure \ref{fig:bang_per_buck}, this plot shows how many repeated transit or eclipse observations with \textit{JWST} are required to detect the presence of a 0.01 bar atmosphere. We compare four different detection methods: transit spectroscopy (without and with clouds), eclipse spectroscopy, eclipse photometry, and phase curves. For phase curves we convert the observation time needed into an equivalent amount of transits or eclipses. In contrast to Figure \ref{fig:bang_per_buck}, the surface pressure here is 0.01 bar so the atmosphere is thin and heat redistribution is relatively inefficient.}
\label{fig:bang_per_buck02}
\end{figure*}

Figure \ref{fig:bang_per_buck} also indicates that transit spectroscopy is promising, especially for cool planets with low surface gravity and lower-MMW atmospheres.
This result is sensitive, however, to the potential presence of clouds or hazes. For example, if GJ1132b had a clear H$_2$O atmosphere, a single \textit{JWST} transit should be able to detect this atmosphere at almost 3$\sigma$ confidence. In contrast, for a cloudy transit spectrum it would take about 10 transits to build up the same detection confidence. Our result is in qualitative agreement with previous theoretical work \citep{fortney2005} as well as observations of hot Jupiters \citep{sing16} which show that transit spectroscopy is highly susceptible to high-altitude clouds and hazes.

Eclipse spectroscopy becomes more promising than transit spectroscopy on hotter planets with higher-MMW atmospheres. For example, even if LHS3844b and GJ1132b had clear CO$_2$ atmospheres, these atmospheres would be easier to detect via eclipse spectroscopy than transit spectroscopy.
{This dependence on temperature is driven by the two methods' different sensitivities: a planet's transit signal is proportional to the atmospheric scale height, which increases linearly with temperature, whereas a planet's eclipse signal is proportional to the Planck function \citep{cowan2015}, which increases much faster than linearly with temperature at the relevant wavelengths.
For example, the peak of the Planck function $\max(B_\lambda)$ for these three planets is inside or close to the MIRI/LRS bandpass, and $\max(B_\lambda) \propto T^5$. As long as the atmospheric MMW remains high, eclipse spectroscopy thus always becomes more favorable on hotter planets.
}

Thermal phase curves are surprisingly attractive when compared to transit and eclipse spectroscopy, even though they require a relatively large observational investment up front.
For example, phase curves always outperform cloudy transit spectroscopy and eclipse spectroscopy in Figure \ref{fig:bang_per_buck}.
This is particularly the case for short-period planets like LHS3844b, where its short orbital period means that a single phase curve is relatively cheap compared to repeated transits or eclipses.

Figure \ref{fig:bang_per_buck} also shows that detection methods with higher spectral resolution generally improve quicker with repeated observations. This effect can be seen for LHS3844b with an H$_2$O atmosphere where the detection probability for a cloudy transit first lags behind, but then rises faster than, eclipse spectroscopy. The underlying reason is that we bin the emission spectrum of an H$_2$O atmosphere down to just two spectral points, whereas the transit spectrum contains four points (see Section \ref{sec:jwst_obs}). Even though the transit spectrum thus starts at a lower SNR than the emission spectrum, it contains more degrees of freedom and its SNR improves faster with more observations.

Figure \ref{fig:bang_per_buck02} is the same as Figure \ref{fig:bang_per_buck}, but shows our results for a thin atmosphere with less efficient heat redistribution. We find that thinner atmospheres are more difficult to detect overall, but transit and eclipse spectroscopy are less affected by low surface pressure than eclipse photometry and thermal phase curves.
For example, reducing the atmosphere's thickness on GJ1132b from 1 bar to 0.01 bar roughly doubles the observing time necessary to detect CO$_2$ spectral features via eclipse spectroscopy. In contrast, the same reduction in atmospheric thickness {on GJ1132b} increases the observational effort for eclipse photometry {by a factor of five and for} phase curves by more than a factor of ten.

The comparison of Figures \ref{fig:bang_per_buck} and \ref{fig:bang_per_buck02} shows that no atmospheric detection method always outperforms all others.
However, as long as planets that are favorable for observations also have moderately thick atmospheres, these atmospheres can be detected with {one to two} eclipses. Eclipse photometry is therefore a promising screening method that can justify and guide \textit{JWST} follow-up efforts.

%%%%%%%%%%%%%%%%%%%%%%%%%%%%
%% --- added ....
\section{Potential for Future Atmospheric Searches}
\label{sec:future}
%%%%%%%%%%%%%%%%%%%%%%%%%%%%

\begin{figure*}
\centering
\includegraphics[width=0.85\linewidth, clip]{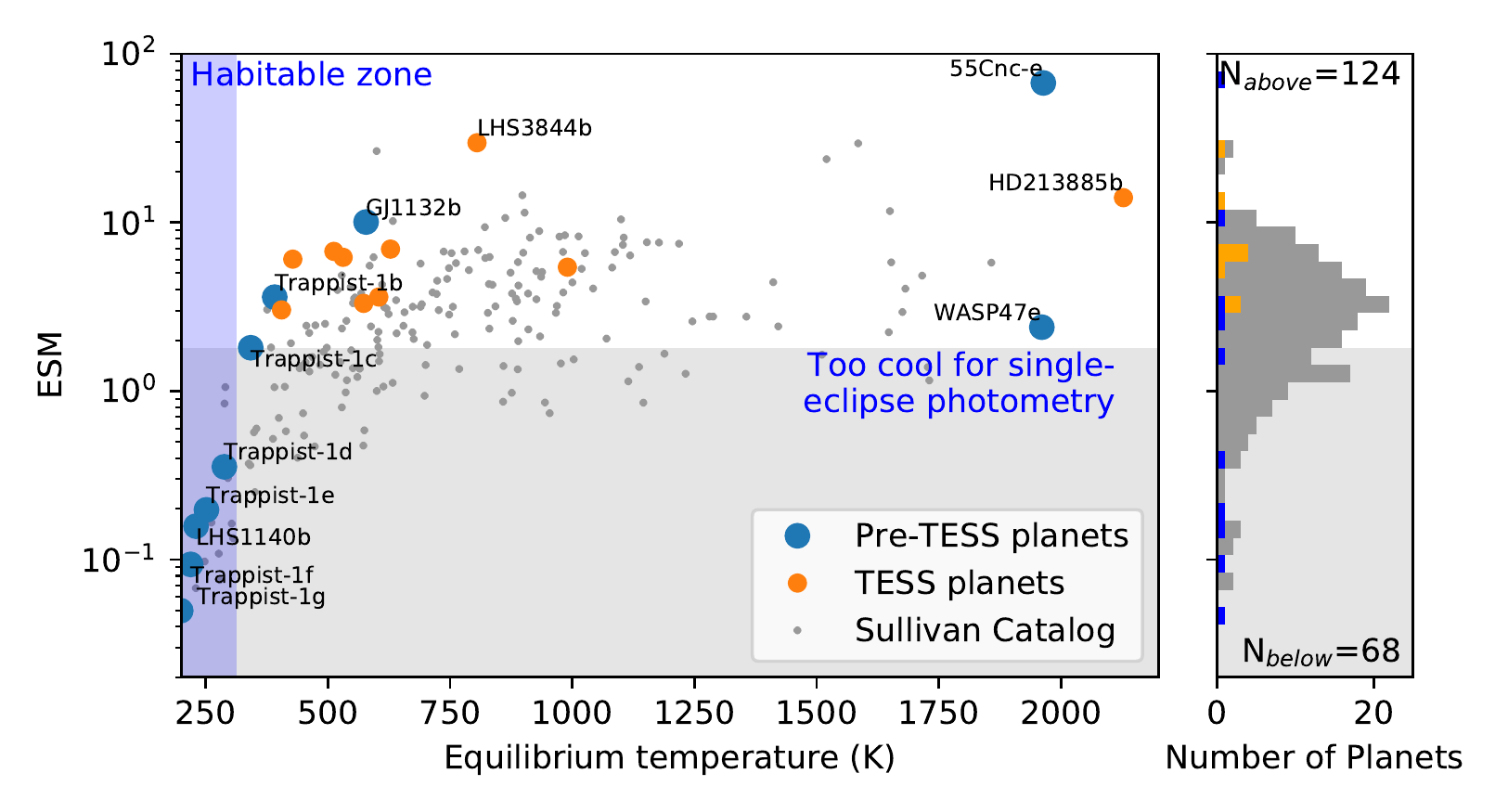}
\caption{
{This plot shows the number of potential rocky planets that are accessible to single-eclipse photometry. Left shows the Emission Spectroscopy Metric (ESM) from \citet{kempton2018} for known planets (blue and orange) as well as simulated \textit{TESS} planets from \citet[][grey circles]{sullivan2015}. We consider all planets with an ESM greater than that of TRAPPIST-1c to be feasible targets (see text). The blue shaded region delineates the habitable zone, the grey shaded region delineates TRAPPIST-1c's ESM. Right shows the histogram of all planets as a function of ESM; there are N$_{above}=124$ planets in the Sullivan catalog with an ESM greater than that of TRAPPIST-1c.
}
}
\label{fig:tess_targets}
\end{figure*}

{Our calculations consider three planets that are widely agreed-upon to be excellent targets for atmospheric characterization with \textit{JWST} \citep{morley2017,louie2018}. However, the number of rocky exoplanets that are potentially suitable for atmospheric characterization is growing rapidly thanks to the \textit{TESS} mission, so how many more planets could \textit{JWST} feasibly search for candidate atmospheres?}

{
To address this question, we perform an estimate using the exoplanet catalog from \citet{sullivan2015} and the analytical emission spectroscopy metric (ESM) from \citet{kempton2018}. Although the ESM is only an analytical approximation, it adequately captures the $\chi^2$ ordering in Figure \ref{fig:eclipse_obs}. LHS3844b has the highest $\chi^2$ value compared to a bare rock and it also has the highest ESM value of 30, while GJ1132b and TRAPPIST-1b have appropriately smaller ESMs of 10 and 4 respectively. We note that these ESM values are slightly different from those reported in \citet{kempton2018} due to different assumed stellar properties; here we use stellar properties that match Table \ref{tab:parameters}.
}

{
We first estimate an ESM threshold below which a single eclipse is no longer sufficient for detecting an atmosphere. To do so we focus on the TRAPPIST-1 system, because all of its planets share the same host star. We rescale the emission spectrum of TRAPPIST-1b with a 1 bar atmosphere from \texttt{HELIOS} to that of the colder TRAPPIST-1 planets, using the ratio of the planets' bare-rock Planck functions as the scaling factor. We find that $\chi^2$ of TRAPPIST-1c's spectrum relative to a bare rock already drops to $0.5$ for a CO$_2$ atmosphere and $4.4$ for a H$_2$O atmosphere. For TRAPPIST-1d this value drops further, to $0.1$ for CO$_2$ and $0.9$ for H$_2$O. We therefore consider TRAPPIST-1c, which has an ESM of $1.8$, to be the marginal case above which single-eclipse photometry can still detect a 1 bar atmosphere of the right composition with some confidence.
}

{
Figure \ref{fig:tess_targets} shows the ESM for all simulated rocky planets from the Sullivan catalog as well as the ESM for a number of actual rocky planets. To narrow down the Sullivan catalog we only consider planets smaller than 1.5 times Earth's radius to be rocky.
We note that the occurrence rates in the Sullivan catalog are likely biased for planets around small host stars, and the number of planets found by \textit{TESS} could be higher depending on the multiplicity of planets around small host stars \citep{louie2018}.
Blue dots in Figure \ref{fig:tess_targets} show some favorable rocky planets detected before the launch of \textit{TESS} \citep{demory2011a,winn2011,berta-thompson2015,becker2015,dittmann2017,gillon2017}, orange dots show planets or planet candidates that were announced recently \citep{vanderspek2019,dumusque2019,gunther2019,kostov2019,espinoza2019,luque2019,crossfield2019,winters2019}, and the blue shaded region indicates the habitable zone \citep{yang2014}.
}

{
According to Figure \ref{fig:tess_targets}, \textit{TESS} should detect $124$ rocky planets that are favorable targets for atmospheric detection via eclipse photometry. Detailed follow-up of these planets will be more difficult, however, as only $19$ of them have an ESM greater than GJ1132b's and only one of them has a transmission spectroscopy metric (TSM) greater than GJ1132b's \citep[also see][]{kempton2018}. For reference, in the last year \textit{TESS} has discovered two planets with an ESM greater than GJ1132b's, namely LHS3844b and HD213885b, and eight planets with an ESM smaller than GJ1132b's (Fig.~\ref{fig:tess_targets}).
}

{
The prospect that \textit{TESS} will find many targets that are amenable to eclipse photometry, but difficult to characterize in more detail, thus favors statistical surveys. For example, theoretical models predict that atmospheric escape is strongly sensitive to host star type via the host star's XUV output \citep{zahnle17}.
Out of the 124 planets from the Sullivan catalog, 20 of them orbit late M dwarfs with stellar temperatures less than 3300 K, while 85 of them orbit mid to early M dwarfs with stellar temperatures between 3300 and 4000 K.
A \textit{JWST} survey could thus empirically test whether there is a strong correlation between host star type and the ability of rocky exoplanets to retain an atmosphere, which is an important constraint for planetary evolution models as well as future astrobiological searches.
}

%%%%%%%%%%%%%%%%%%%%%%%%%%%%
\section{Discussion}
\label{sec:discussion}
%%%%%%%%%%%%%%%%%%%%%%%%%%%%

%% --- Discuss false positives here.
\subsection{False Positives}

Bare rocks with high Bond albedos are an important false positive scenario for our proposal because, just like a thick atmosphere, a high albedo would also reduce a planet's dayside thermal emission (see Eqn.~\ref{eqn:daybudget}).

We consider this false positive scenario unlikely.
First, in a companion paper we compile geological and laboratory evidence which suggests that the surface albedos of rocky exoplanets with equilibrium temperatures in the range of the planets we consider here, $300$ K$<T_{eq}<880$ K, should be low (Mansfield et al, submitted). The underlying reason is that many geologic surfaces with high albedo (e.g., granites, clays) are either water-rich or require liquid water to form, which is unlikely for planets with $T_{eq} > 300$ K because these planets are located inside the inner edge of the M dwarf habitable zone. At the same time, planets with $T_{eq} > 880$ K are hot enough to vaporize substantial amounts of rock on their dayside over geologic timescales. This partial vaporization would preferentially remove more volatile species, and so could leave behind a low-volatile residue rich in Aluminium and Calcium compounds that have high albedos. By focusing on planets with $300$ K$\leq T_{eq}\leq 880$ K, we minimize the possibility of either false positive scenario occuring.

Second, Solar System analogs similarly suggest that exoplanets without atmospheres will have low albedos \citep{madden2018}. 
Notable bare rocks in the Solar system include Mercury, which has an albedo of less than 0.1, the Moon and Ceres, which have albedos of 0.1-0.15, and asteroids, the majority of which have an albedo less than 0.2 \citep{wright2016}.
There are some airless bodies in the Solar system with high albedos, such as Europa with an albedo of $\sim 0.6$, and Jupiter's moon Io with an albedo of $\sim 0.5$. However, neither Europa nor Io are plausible analogs for short-period exoplanets, because their high albedos are caused by water ice and condensed sulfur species that are unstable inside the inner edge of the habitable zone. {We note that sulfur can exist in liquid form inside the inner edge of the habitable zone \citep{theilig1982}, but any sulfur pools or oceans would again have a low albedo \citep{nelson1983}.}
Based on these considerations, we believe it is justified to assume that other rocky exoplanets will have similarly low surface albedos.

Another false positive scenario is a planet that is not tidally locked. In this case the dayside emission temperature would be lower due to the planet's rotation instead of atmospheric heat redistribution.
Based on theoretical arguments we consider this scenario unlikely. First, atmospheric models show that even non-synchronous rotators have day-night temperature contrasts similar to tidally locked rotators if the planet is sufficiently hot and the atmosphere sufficiently thin, so that the atmosphere's radiative timescale is short compared to the planet's rotation period \citep{rauscher2014}.
Following this argument GJ1132b in a 3:2 spin-orbit resonance would appear effectively tidally locked if its atmosphere were thinner than $\sim$0.5 bar. If the dayside temperature were then observed to be much cooler than a tidally locked bare rock, this would still indicate a relatively thick atmosphere.
Second, tidal models suggest that non-synchronous rotation is unlikely for short-period planets orbiting small host stars \citep{leconte2015,barnes2017}, which includes all three targets we consider above, even though it might become relevant for planets around late K and early M dwarfs that are located inside their host stars' habitable zone. 

A final false positive scenario is dynamical heat redistribution by a lava ocean instead of an atmosphere. This scenario does not apply to planets like LHS3844b or GJ1132b, and is only feasible on planets like 55 Cancri e which are hot enough that their dayside is molten while simultaneously cool enough that rock vapor does not form a thick atmosphere. However, even for 55 Cancri e we do not consider heat redistribution by a lava ocean likely based on theoretical estimates that lava ocean currents are too slow to affect planetary heat redistribution in the absence of a wind-driven circulation \citep{kite2016b}.

%% --- Discuss false negatives here.
\subsection{False Negatives}

Thin atmospheres with inefficient heat redistribution are a
likely false negative scenario for our proposal. 
Such atmospheres would be easier to detect via transit and eclipse spectroscopy, or potentially by detecting the planet's nightside emission via thermal phase curves (Fig.~\ref{fig:bang_per_buck02}).

We note that some atmospheres might be thin but still have significant cloud cover, analogous to how Mars' atmosphere is thin but can produce reflective clouds as well as optically thick planet-encircling dust storms. Such thin atmospheres might not reveal their presence via the atmosphere's heat redistribution but could still be detectable in eclipse photometry through the clouds' effect on the planet's albedo, which is a possibility for identifying candidate atmospheres that we explore in a companion paper (Mansfield et al, submitted).

%% --- Discuss ....
\subsection{Additional physics}

Our models do not include the impact of clouds on the dayside emission spectrum, nor do we consider the increased day-night latent heat transport in atmospheres with condensation.
Both processes should tend to reduce the dayside brightness temperature, and thus could affect the quantitative interpretation of eclipse observations. However, given that both processes are atmospheric phenomena, a low observed brightness temperature would thus still be indicative of an atmosphere.

We also assume blackbody spectra for the planet surface, even though minerals can induce spectral features on airless bodies \citep{hu2012}.
Surface-induced spectral features are an important potential false positive for eclipse spectroscopy, which should be explored in future work.
Nevertheless, we don't expect that surface spectral features would negatively affect atmospheric detection via eclipse photometry. The underlying reason is that the emissivity of many minerals tends to increase from shorter to longer wavelengths, so the brightness temperature an observer sees at relatively long wavelengths in the MIRI bandpass will be biased high (Mansfield et al, submitted). This bias works in the opposite direction of atmospheric heat transport, so an observed cool dayside would be even more indicative of an atmosphere if we accounted for surface spectral features than it is with a blackbody surface.

%%%%%%%%%%%%%%%%%%%%%%%%%%%%
\section{Conclusions}
\label{sec:conclusions}
%%%%%%%%%%%%%%%%%%%%%%%%%%%%

%% --- Conclusions here.

We have used simulated transit spectra, eclipse spectra, and \textit{JWST} noise calculations to compare the efficiency of different methods for detecting atmospheres on rocky exoplanets. Focusing on three planets that are high-priority targets for atmospheric characterization with \textit{JWST}, we find the following:
\begin{enumerate}
    \item For targets that are amenable to atmospheric follow-up, {one to two} eclipses with \textit{JWST} should be sufficient to detect the heat redistribution signal of a moderately thick atmosphere with $\mathcal{O}(1)$ bar of surface pressure. Eclipse photometry is therefore a promising method for quickly identifying candidate atmospheres.
    \item Candidate atmospheres can be confirmed by follow-up transit spectroscopy, eclipse spectroscopy, or thermal phase curves. No follow-up technique is always superior, and the best observational strategy will depend on stellar, planetary, and atmospheric parameters (Figures \ref{fig:bang_per_buck}, \ref{fig:bang_per_buck02}).
    In particular, if rocky exoplanet atmospheres are cloud- and haze-free, transit spectroscopy will be attractive for a broad range of targets.
    If transit spectroscopy is muted by hazes, eclipse spectroscopy and thermal phase curves might still be viable techniques for atmospheric characterization.
\end{enumerate}

{
In addition, we have estimated how many rocky exoplanets will be detected by \textit{TESS} that could be studied using eclipse photometry on \textit{JWST}. \textit{TESS} will find more than $100$ hot, non-habitable planets that are potentially amenable to this technique (Figure~\ref{fig:tess_targets}). About $10$ such planets have already been announced over the past year.}
{
A comparatively modest \textit{JWST} Large program (i.e., $>$\,75\,hours) should be sufficient to screen the most accessible of these planets for candidate atmospheres, and would then provide a stepping stone to more comprehensive follow-up campaigns.
}
{Eclipse photometry} is also an attractive option for future statistical surveys to constrain what fraction {of rocky planets} host thick atmospheres, which is an important unknown in the search for life around other stars.
%

%%%%%%%%%%%%%%%%%%%%%%%%%%%%
\acknowledgments

{We thank an anonymous reviewer for positive and constructive feedback that helped improve this work.}
D.D.B.K. was supported by a James McDonnell Foundation postdoctoral fellowship.
M.~Malik acknowledges support from the Swiss National Science Foundation under the Early Postdoc Mobility grant P2BEP2\_181705.
J.L.B. acknowledges support from the David and Lucile Packard Foundation.
E.~Kite was supported by NASA grant NNX16AB44G.
E.~Kempton acknowledges support from the National Science Foundation under Grant No.1654295 and from the Research Corporation for Science Advancement through their Cottrell Scholar program.
This work was supported by the NASA Astrobiology Program Grant Number
80NSSC18K0829 and benefited from participation in the NASA Nexus for
Exoplanet Systems Science research coordination network.
%

%%%%%%%%%%%%%%%%%%%%%%%%%%%%%

\bibliography{ZoteroLibrary-latest,new,matej_bib}

\end{document}